\documentclass{article}%
\usepackage{amsmath}
\usepackage{amsfonts}
\usepackage{amssymb}
\usepackage{graphicx}%
\providecommand{\U}[1]{\protect\rule{.1in}{.1in}}

\begin{document}

\author{Giuseppe Castagnoli\\Elsag Bailey ICT Division and Quantum Information Laboratory\thanks{retired} }
\title{Highlighting the mechanism of the quantum speedup by time-symmetric and
relational quantum mechanics}
\maketitle

\begin{abstract}
Bob hides a ball in one of four drawers. Alice is to locate it. Classically
she has to open up to three drawers, quantally just one. The fundamental
reason for this \textit{quantum speedup} is not known. The usual
representation of the quantum algorithm is limited to the process of solving
the problem. We extend it to the process of setting the problem. The number of
the drawer with the ball becomes a unitary transformation of the random
outcome of the preparation measurement. This extended, time-symmetric,
representation brings in relational quantum mechanics. It is \textit{with
respect to} Bob and any external observer and cannot be with respect to Alice.
It would tell her the number of the drawer with the ball before she opens any
drawer. To Alice, the projection of the quantum state due to the preparation
measurement should be retarded at the end of her search; in the input state of
the search, the drawer number is determined to Bob and undetermined to Alice.
We show that, mathematically, one can ascribe any part of the selection of the
random outcome of the preparation measurement to the final Alice's
measurement. Ascribing half of it explains the speedup of the present
algorithm. This projects the input state to Alice on a state of lower entropy
where she knows half of the number of the drawer with the ball in advance. The
quantum algorithm turns out to be a sum over histories in each of which Alice
knows in advance that the ball is in a pair of drawers and locates it by
opening one of the two. In the sample of quantum algorithms examined, the part
of the random outcome of the initial measurement selected by the final
measurement is one half or slightly above it. Conversely, given an oracle
problem, the assumption it is one half always corresponds to an existing
quantum algorithm and gives the order of magnitude of the number of oracle
queries required by the optimal one.

\end{abstract}

\section{Foreword}

A quantum algorithm is said to deliver a quantum speedup when it requires
fewer computation steps than its classical counterpart, sometimes fewer than
the number demonstrably necessary in the classical case. An example is as
follows. Bob, the problem setter, hides a ball in one of four drawers, Alice,
the problem solver, is to locate it. Classically, Alice might have to open up
to three drawers. With Grover quantum search algorithm $\left[  1\right]  $,
it always takes one. More in general, given $N=2^{n}$ drawers, this algorithm
locates the ball by opening $\operatorname*{O}\left(  2^{n/2}\right)  $
drawers, against the $\operatorname*{O}\left(  2^{n}\right)  $\ of the
classical case. It yields a "quadratic" speedup.

The drawers and ball problem is an example of oracle problem and the operation
of opening a drawer and checking whether the ball is in it is an example of
oracle query. More in general, the speedup comes from comparing the number of
oracle queries required to solve certain oracle problems in a quantum and
classical way.

Its violating the limits applying to classical time-evolutions relates the
speedup to the violation of the temporal Bell inequality of Leggett and Garg
$\left[  2\right]  $, the information-theoretic one of Braunstein and Caves
$\left[  3\right]  $ and, particularly, the one formulated by Morikoshi
$\left[  4\right]  $ exactly in the case of Grover algorithm.

While there is an important body of literature on various mathematical
relations between speedup and other quantum features, like entanglement and
discord, an explanation of the fundamental physical mechanism of the speedup
has never been pinpointed.

Whether there is such an explanation is controversial. From the one side,
since the speedup is a mathematical property of unitary transformations, it
can be studied independently of the foundations of quantum mechanics. From the
other, the very notion of quantum computation and speedup was born out of the
assumption that computation is a physical thing, not only a mathematical
abstraction, see the seminal papers $\left[  5\div9\right]  $. The natural
premise of this work is that there may be a utility in looking at the speedup
from a fundamental physical perspective.

\section{Introduction}

The present investigation of the foundations of the speedup begins by
completing the physical representation of the problem-solving process. The
usual representation of quantum algorithms is limited to the process of
solving the problem. We extend it to the process of setting the problem --\ to
the preparation of the number of the drawer with the ball in the case of
Grover algorithm. To this end, to the usual quantum registers that contain the
number of the drawer that Alice wants to open and the result of opening it, we
should add an imaginary register that contains the number of the drawer with
the ball.

For reasons that will soon become clear, we assume that the initial state of
the imaginary register is a uniform mixture of all the possible numbers of the
drawer with the ball. At time $t_{0}$, Bob measures the content of this
register selecting a number at random, then he unitarily transforms it into
the desired number, at time $t_{1}$. Alice, by opening drawers, unitarily
transforms the state of the quantum registers at time $t_{1}$ into a tensor
product of the number of the drawer with the ball chosen by Bob and the
corresponding solution of the problem (that same number but in another
register), at time $t_{2}$. Then she reads the solution by a final measurement.

This extended representation implies a further extension. Preparing the
imaginary register with the number of the drawer with the ball amounts to
preparing it in a sharp state. This cannot be the representation of the
quantum state with respect to the observer Alice. It would tell her the number
of the drawer with the ball before she opens any drawer. As well known, to
Alice this number must be hidden inside a black box.

We represent this physically by resorting to relational quantum mechanics
$\left[  13\right]  $, where quantum states are observer dependent. A state
can be sharp to an observer and a quantum superposition, or a mixture, to
another observer.

We represent the quantum algorithm to Alice by retarding the projection of the
quantum state induced by the initial Bob's measurement at the end of the
unitary part of her problem solving action -- as well known the projection
induced by a quantum measurement can be retarded or advanced at will along a
unitary transformation that respectively follows or precedes the measurement.

With this, in the representation to Alice, the state of the imaginary register
at time $t_{1}$, after Bob's preparation of the number of the drawer with the
ball, is still a mixture of all the possible numbers. Its two bit entropy
represents Alice's complete ignorance of the number in question. By opening
drawers, Alice unitarily transforms this state into a mixture of tensor
products, each a number of the drawer with the ball and the corresponding
solution, at time $t_{2}$. The entropy of the solution, the trace over the
drawer numbers, is also two bits.

This representation of the quantum algorithm is time-symmetric $\left[
10\right]  $. The two bit entropy of the solution is zeroed by either the
retarded projection induced by the initial Bob's measurement or the projection
induced by the final Alice's measurement. Since the two projections can occur
in any order or simultaneously, there is an ambiguity about which measurement
zeroes the entropy of the solution.

It is useful to recall the tenet of time-symmetric quantum mechanics we will
resort to, namely that the maximal description of the quantum process --
initial measurement at time $t_{0}$, unitary transformation, and final
measurement at time $t_{2}$ -- requires knowledge not only of the outcome of
the initial measurement, also of that of the final measurement. This implies
that the latter outcome has back in time implications on the process at times
$t<t_{2}$ see also $\left[  11,12\right]  $.

The present work is an exploration of the assumption that the zeroing of the
entropy of the solution shares between the two measurements. For Occam
razor\footnote{In Newton's formulation, it states \textquotedblleft\textit{We
are to admit no more causes of natural things than such that are both true and
sufficient to explain their appearances}\textquotedblright\ $\left[
14\right]  $.}, we exclude any redundancy between the two measurements. We
assume that, when the two measurements are considered together, they reduce to
partial measurements that contribute in a complementary and non-redundant way
to the zeroing of the entropy of the solution. By this we mean that no
information about the solution provided by one partial measurement is provided
by the other.

To reconstruct the selections performed by the complete measurements, the
projection of the quantum state induced by Bob's partial measurement must
propagate forward in time, from $t_{0}$ to $t_{2}$, along the unitary
transformation comprised between the two measurements, that induced by Alice's
partial measurement backward in time, from $t_{2}$ to $t_{0}$, along the
inverse transformation.

The backward propagation is a quantum feedback. Let $\mathcal{R}$\ be the
fraction of entropy of the solution whose reduction is ascribed to Alice's
measurement. $\mathcal{R}=0$ means that the zeroing of the entropy is all
ascribed to Bob's measurement, $\mathcal{R}=1$ all to Alice's measurement,
$\mathcal{R}=\frac{1}{2}$\ that it is equally shared between the two
measurements. $\mathcal{R}$ is a sort of retrocausality index.

We will see that, at time $t_{0}$, the backward propagation selects the
$\mathcal{R}$-th part of the information that specifies the random outcome of
Bob's measurement. At time $t_{1}$, it projects the input state of the quantum
algorithm to Alice, one of complete ignorance of the number of the drawer with
the ball chosen by Bob, on a state of lower entropy where she knows the
$\mathcal{R}$-th part of the number in question (of the information that
specifies it) in advance, before opening any drawer. Alice uses this advanced
knowledge to locate the ball opening fewer drawers.

The speedup of the four drawers case is quantitatively justified by the
assumption that $\mathcal{R}=\frac{1}{2}$. Consequently Alice knows in advance
half of the information that specifies the number of the drawer with the ball,
namely half of the solution she will read in the future. She knows that the
ball is in a pair of drawers instead of one of four. There are of course
several ways of taking a pair of drawers such that one has the ball in it. The
quantum algorithm turns out to be a sum over classical histories in each of
which Alice knows in advance that the ball is in one of such pairs ad locates
it by opening either drawer.

The present explanation of the speedup, given any oracle problem and a value
of $\mathcal{R}$, provides the corresponding number of oracle queries required
to solve the problem quantally. Conversely, given a quantum algorithm, it
provides the value of $\mathcal{R}$\ that explains its speedup.

We have ascertained the value of $\mathcal{R}$\ that explains the speedup of
the major quantum algorithms. Beyond Grover algorithm with $n=2$, we have
$\mathcal{R}=\frac{1}{2}$ in the algorithm of Deutsch\&Jozsa and that of Simon
-- consequently in the quantum algorithms of the Abelian hidden subgroup --
which noticeably comprise Shor's factorization algorithm. All these quantum
algorithms are optimal and provide the solution of the problem with a single
oracle query.

We should make a clarification. When $n>2$, Grover algorithm ceases to provide
the solution with absolute certainty. We should resort to the revisitation of
this algorithm due to Long $\left[  15,16\right]  $. Grover/Long algorithm can
be tuned to provide the solution with certainty on any number of oracle
queries (drawer openings) provided it is above the minimum number required by
the optimal quantum algorithm, which is about $\frac{\pi}{4}2^{n/2}$. By the
way, this is also the number required by the original Grover
algorithm\footnote{Reference $\left[  17\right]  $\ showed that Grover
algorithm is optimal in the order of magnitude. Reference $\left[  16\right]
$, posterior of 16 years, shows that Grover/Long algorithm is exactly
optimal.} (which does not provide the solution with absolute certainty however).

When $n$ goes past $2$, the speedup of Grover/Long algorithm is explained by
values of $\mathcal{R}$\ slightly above $\frac{1}{2}$, which go back to
$\frac{1}{2}$\ for $n\rightarrow\infty$. We should note that $\mathcal{R}%
=\frac{1}{2}$ anyway corresponds to an existing quantum algorithm, suboptimal
for $n>2$. This is in fact Long's algorithm tuned on the $2^{n/2}%
-1\simeq2^{n/2}$ queries required when $\mathcal{R}=\frac{1}{2}$.

Summing up, in all the cases examined, the number of oracle queries required
in the assumption that $\mathcal{R}=\frac{1}{2}$ is always that of an existing
quantum algorithm and a good approximation of the number required by the
optimal one. If this were true in general, we would have an important result.
The problem of quantum query complexity, namely of the number of oracle
queries required to solve a generic oracle problem in an optimal quantum way,
is open.

This seems to be an interesting prospect opened by the present retrocausal
analysis of the speedup. Further study could check the value of $\mathcal{R}$
for other classes of quantum algorithms, whether $\mathcal{R}=\frac{1}{2}$\ is
always attainable, whether it is reasonable to think it always provides the
order of magnitude of the number of oracle queries required by the optimal
quantum algorithm. The present work is an exploration.

The work belongs to an evolutionary approach $\left[  18\div20\right]  $.
Besides improving the clarity and the precision of our analysis of the
speedup, this time we have provided a procedure for calculating the number of
queries required to solve a generic oracle problem under the $\mathcal{R}%
=\frac{1}{2}$ assumption.

The work has points of contact with works of Morikoshi. In $\left[  21\right]
$, this author highlights the problem-solution symmetry of Grover's and the
phase estimation algorithms and notes it may be relevant for the explanation
of the speedup. In $\left[  4\right]  $, he shows that Grover algorithm
violates a temporal Bell inequality. The present explanation of the speedup
might establish a connection between this violation and quantum
retrocausality, in the form of quantum feedback from the final to the initial measurement.

The form of quantum retrocausality utilized in the present work has been
inspired by the work of Dolev and Elitzur $\left[  22\right]  $ on the
non-sequential behavior of the wave function highlighted by partial measurement.

\section{Quantum feedback}

We formalize the relation between quantum speedup and retrocausality, first in
Grover algorithm then in the general case of oracle quantum computing.

\subsection{Grover algorithm}

\subsubsection{Time-symmetric, relativized representations}

Let $\mathbf{b}$ and $\mathbf{a}$, belonging to $\left\{  0,1\right\}  ^{n}$,
be respectively the number of the drawer with the ball and that of the drawer
that Alice wants to open. Opening drawer $\mathbf{a}$ amounts to evaluating
the function $f_{\mathbf{b}}\left(  \mathbf{a}\right)  $, which is one if
$\mathbf{a=b}$ and zero otherwise -- tells Alice whether the ball is in drawer
$\mathbf{a}$.

Bob selects one of the functions $f_{\mathbf{b}}\left(  \mathbf{a}\right)  $
(ie a value of $\mathbf{b}$) and gives Alice the black box that computes it.
Alice knows that the function computed by the black box is one of the
$f_{\mathbf{b}}\left(  \mathbf{a}\right)  $'s, but does not know which one.
She is to find the value of $\mathbf{b}$ (ie a \textit{characteristic} of the
function computed by the black box) by performing function evaluations (ie
oracle queries) for appropriate values of $\mathbf{a}$. Of course Alice is
forbidden to inspect the suffix of the function contained in the black box
(hence the name).

The usual representation of Grover's algorithm uses two quantum registers.
Alice's register $A$ contains the argument of the function $\mathbf{a}$.
Register $V$ is meant to contain the result of the computation of
$f_{\mathbf{b}}\left(  \mathbf{a}\right)  $, modulo 2 added to its former
content for logical reversibility. To physically represent also the problem
setting, we need an imaginary register $B$ that contains $\mathbf{b}$.\ 

The basis vectors of registers $A$ and $B$ are respectively $\left\vert
00\right\rangle _{A},\left\vert 01\right\rangle _{A},\left\vert
10\right\rangle _{A},\left\vert 11\right\rangle _{A}$ and $\left\vert
00\right\rangle _{B},\left\vert 01\right\rangle _{B},\left\vert
10\right\rangle _{B},\left\vert 11\right\rangle _{B}$, where $00,01,~...$ are
the drawer numbers in binary notation. Those of register $V$ are $\left\vert
0\right\rangle _{V}$ and $\left\vert 1\right\rangle _{V}$. We assume that,
initially, register $B$ is in a maximally mixed state. Its reduced density
operator is thus:%
\begin{equation}
\rho_{B}=\frac{1}{4}\left(  \left\vert 00\right\rangle _{B}\left\langle
00\right\vert _{B}+\left\vert 01\right\rangle _{B}\left\langle 01\right\vert
_{B}+\left\vert 10\right\rangle _{B}\left\langle 10\right\vert _{B}+\left\vert
11\right\rangle _{B}\left\langle 11\right\vert _{B}\right)  , \label{den}%
\end{equation}
The reason for this assumption will soon become clear.

In view of the sum over histories representation, we need to represent all
states as ket vectors, not matrices. To this end we move to the random phase
representation $\left[  23\right]  $ of state (\ref{den}):%

\begin{equation}
\left\vert \psi\right\rangle _{B}=\frac{1}{2}\left(  \operatorname{e}%
^{i\varphi_{0}}\left\vert 00\right\rangle _{B}+\operatorname{e}^{i\varphi_{1}%
}\left\vert 01\right\rangle _{B}+\operatorname{e}^{i\varphi_{2}}\left\vert
10\right\rangle _{B}+\operatorname{e}^{i\varphi_{3}}\left\vert 11\right\rangle
_{B}\right)  , \label{int}%
\end{equation}
where the $\varphi_{i}$ are independent random phases with uniform
distribution in $\left[  0,2\pi\right]  $. Reading state (\ref{int}) is
simple. It is a mixture of pure states with the phases $\varphi_{0}$,
$\varphi_{1}$, $\varphi_{2}$, $\varphi_{3}$ all different, in fact a dephased
quantum superposition.$\ $The von Neumann entropy of state (\ref{int}) is two
bit. Note that the usual density operator representation is the average over
all $\varphi_{i}$ of the product of the ket by the bra: $\rho_{B}=\left\langle
\left\vert \psi\right\rangle _{B}\left\langle \psi\right\vert _{B}%
\right\rangle _{\forall\varphi_{i}}$.

The overall initial state of the three registers, at time $t_{0}$, is:%
\begin{equation}
\left\vert \psi\right\rangle =\frac{1}{2\sqrt{2}}\left(  \operatorname{e}%
^{i\varphi_{0}}\left\vert 00\right\rangle _{B}+\operatorname{e}^{i\varphi_{1}%
}\left\vert 01\right\rangle _{B}+\operatorname{e}^{i\varphi_{2}}\left\vert
10\right\rangle _{B}+\operatorname{e}^{i\varphi_{3}}\left\vert 11\right\rangle
_{B}\right)  \left\vert 00\right\rangle _{A}\left(  \left\vert 0\right\rangle
_{V}-\left\vert 1\right\rangle _{V}\right)  . \label{in}%
\end{equation}

In order to prepare the problem setting, at time\ $t_{0}$ Bob measures the
\textit{content} of register $B$, namely the observable $\hat{B}$ of
eigenstates the basis vectors $\left\vert 00\right\rangle _{B},\left\vert
01\right\rangle _{B},...$ and eigenvalues respectively $00,01,...$. The
measurement outcome is completely random. Say it comes out the drawer number
$10$. The state immediately after measurement is:%

\begin{equation}
P_{B}\left\vert \psi\right\rangle =\frac{1}{\sqrt{2}}\left\vert
10\right\rangle _{B}\left\vert 00\right\rangle _{A}\left(  \left\vert
0\right\rangle _{V}-\left\vert 1\right\rangle _{V}\right)  , \label{random}%
\end{equation}
where $P_{B}$ is the projection of the quantum state induced by Bob's
measurement. Then Bob applies to register $B$ a unitary transformation $U_{B}$
that changes the random measurement outcome into the desired problem setting,
say $01$. At time $t_{1}$ we will have:
\begin{equation}
U_{B}P_{B}\left\vert \psi\right\rangle =\frac{1}{\sqrt{2}}\left\vert
01\right\rangle _{B}\left\vert 00\right\rangle _{A}\left(  \left\vert
0\right\rangle _{V}-\left\vert 1\right\rangle _{V}\right)  . \label{inbob}%
\end{equation}

State (\ref{random}) is the input state of the quantum algorithm in the
representation extended to the process of setting the problem, which is the
usual representation up to the ket $\left\vert 01\right\rangle _{B}$.

We can see that this extension immediately calls for another one, this time
concerning the actors (observers) on the stage. We have to resort to the
relational quantum mechanics of Rovelli $\left[  13\right]  $. States
(\ref{in}) through (\ref{random}) constitute the description of the initial
part of the quantum algorithm with respect to Bob, the problem setter, and any
other observer who does not act on the problem solving process. They cannot be
the description with respect to Alice, the problem solver. State
(\ref{random}), with register $B$\ in the sharp state $\left\vert
01\right\rangle _{B}$, would tell her, before she opens any drawer, that the
ball is in drawer $01$. Of course the number of the drawer with the ball
should be hidden to Alice -- to her it is inside the black box.

To physically represent this fact, it suffices to retard the projection
$P_{B}$\ until the end of the unitary part of Alice's action. This yields the
representation of the quantum algorithm with respect to Alice.

To her, the state of register $B$ in the input state of the quantum algorithm
is still maximally mixed. In fact, once removed the projection $P_{B}$, the
unitary transformation $U_{B}$ leaves the maximally mixed state of register
$B$ unaltered up to an irrelevant permutation of the random phases. Thus,
disregarding the permutation, state (\ref{in}) is the input state to Alice.

We started with register $B$ in a maximally mixed state to represent the fact
that, to Alice, the problem setting is physically hidden. The two bit entropy
of state (\ref{in}) represents Alice's complete ignorance of the setting.

For completeness, we provide a detailed description of the unitary part of
Alice's problem-solving action. The reader familiar with Grover algorithm can
look just at the input and output state of it.

We have seen that the input state to Alice is%
\begin{equation}
U_{B}\left\vert \psi\right\rangle =\left\vert \psi\right\rangle =\frac
{1}{2\sqrt{2}}\left(  \operatorname{e}^{i\varphi_{0}}\left\vert
00\right\rangle _{B}+\operatorname{e}^{i\varphi_{1}}\left\vert 01\right\rangle
_{B}+\operatorname{e}^{i\varphi_{2}}\left\vert 10\right\rangle _{B}%
+\operatorname{e}^{i\varphi_{3}}\left\vert 11\right\rangle _{B}\right)
\left\vert 00\right\rangle _{A}\left(  \left\vert 0\right\rangle
_{V}-\left\vert 1\right\rangle _{V}\right)  . \label{ina}%
\end{equation}

Alice's action starts with the application of the Hadamard transform $H_{A}$
to register $A$:%
\begin{align}
H_{A}U_{B}\left\vert \psi\right\rangle  &  =\frac{1}{4\sqrt{2}}\left(
\operatorname{e}^{i\varphi_{0}}\left\vert 00\right\rangle _{B}%
+\operatorname{e}^{i\varphi_{1}}\left\vert 01\right\rangle _{B}%
+\operatorname{e}^{i\varphi_{2}}\left\vert 10\right\rangle _{B}%
+\operatorname{e}^{i\varphi_{3}}\left\vert 11\right\rangle _{B}\right)
\label{prein}\\
&  \left(  \left\vert 00\right\rangle _{A}+\left\vert 01\right\rangle
_{A}+\left\vert 10\right\rangle _{A}+\left\vert 11\right\rangle _{A}\right)
\left(  \left\vert 0\right\rangle _{V}-\left\vert 1\right\rangle _{V}\right)
.\nonumber
\end{align}

Function evaluation, performed in quantum parallelism for all the possible
values of the argument, is represented by the unitary transformation $U_{f}$:%
\begin{equation}
U_{f}H_{A}U_{B}\left\vert \psi\right\rangle =\frac{1}{4\sqrt{2}}\left[
\begin{array}
[c]{c}%
\operatorname{e}^{i\varphi_{0}}\left\vert 00\right\rangle _{B}\left(
-\left\vert 00\right\rangle _{A}+\left\vert 01\right\rangle _{A}+\left\vert
10\right\rangle _{A}+\left\vert 11\right\rangle _{A}\right)  +\\
\operatorname{e}^{i\varphi_{1}}\left\vert 01\right\rangle _{B}\left(
\left\vert 00\right\rangle _{A}-\left\vert 01\right\rangle _{A}+\left\vert
10\right\rangle _{A}+\left\vert 11\right\rangle _{A}\right)  +\\
\operatorname{e}^{i\varphi_{2}}\left\vert 10\right\rangle _{B}\left(
\left\vert 00\right\rangle _{A}+\left\vert 01\right\rangle _{A}-\left\vert
10\right\rangle _{A}+\left\vert 11\right\rangle _{A}\right)  +\\
\operatorname{e}^{i\varphi_{3}}\left\vert 11\right\rangle _{B}\left(
\left\vert 00\right\rangle _{A}+\left\vert 01\right\rangle _{A}+\left\vert
10\right\rangle _{A}-\left\vert 11\right\rangle _{A}\right)
\end{array}
\right]  (\left\vert 0\right\rangle _{V}-\left\vert 1\right\rangle _{V}).
\label{second}%
\end{equation}
Note that nothing changes when $f_{\mathbf{b}}\left(  \mathbf{a}\right)  =0$;
when $f_{\mathbf{b}}\left(  \mathbf{a}\right)  =1$ the term $(\left\vert
0\right\rangle _{V}-\left\vert 1\right\rangle _{V})$ changes into $(\left\vert
1\right\rangle _{V}-\left\vert 0\right\rangle _{V})=-(\left\vert
0\right\rangle _{V}-\left\vert 1\right\rangle _{V})$, the minus sign then goes
to the appropriate basis vector of register $A$.\ 

In (\ref{second}), four orthogonal states of register $B$, each a different
value of $\mathbf{b}$, one by one multiply four orthogonal states of register
$A$. This means that the information that specifies the value of $\mathbf{b}$
has propagated to register $A$.

To make this information accessible to measurement, Alice applies to register
$A$\ the unitary transformation $\Im_{A}$ (the \textit{inversion about the
mean}), which diagonalizes the reduced density operator of register $A$
yielding the output state:%
\begin{align}
\Im_{A}U_{f}H_{A}U_{B}\left\vert \psi\right\rangle  &  =\frac{1}{2\sqrt{2}%
}\left(  \operatorname{e}^{i\varphi_{0}}\left\vert 00\right\rangle
_{B}\left\vert 00\right\rangle _{A}+\operatorname{e}^{i\varphi_{1}}\left\vert
01\right\rangle _{B}\left\vert 01\right\rangle _{A}+\operatorname{e}%
^{i\varphi_{2}}\left\vert 10\right\rangle _{B}\left\vert 10\right\rangle
_{A}+\operatorname{e}^{i\varphi_{3}}\left\vert 11\right\rangle _{B}\left\vert
11\right\rangle _{A}\right) \nonumber\\
&  \left(  \left\vert 0\right\rangle _{V}-\left\vert 1\right\rangle
_{V}\right)  . \label{outputa}%
\end{align}

We can see that, for each possible problem setting, Alice has reconstructed it
in register $A$. Eventually, at time $t_{2}$, she acquires the reconstruction
by reading the content of register $A$. In other words, she measures the
observable $\hat{A}$, of eigenstates $\left\vert 00\right\rangle
_{A},\left\vert 01\right\rangle _{A},...$ and eigenvalues respectively
$00,01,...$ -- note that $\hat{A}$ commutes with $\hat{B}$. We should keep in
mind that the output state (\ref{outputa}) is relativized to Alice. The output
state to Bob and any other observer is
\begin{equation}
\Im_{A}U_{f}H_{A}U_{B}P_{B}\left\vert \psi\right\rangle =\frac{1}{\sqrt{2}%
}\left\vert 01\right\rangle _{B}\left\vert 01\right\rangle _{A}\left(
\left\vert 0\right\rangle _{V}-\left\vert 1\right\rangle _{V}\right)  .
\label{outputb}%
\end{equation}
The outcome of measuring $\hat{A}$\ in state (\ref{outputa}) is unpredictable
to Alice, it is already $01$ to any other observer. Also to Alice, the state
immediately after measurement is:%
\begin{equation}
P_{A}\Im_{A}U_{f}H_{A}U_{B}\left\vert \psi\right\rangle =\frac{1}{\sqrt{2}%
}\left\vert 01\right\rangle _{B}\left\vert 01\right\rangle _{A}\left(
\left\vert 0\right\rangle _{V}-\left\vert 1\right\rangle _{V}\right)  .
\label{alice}%
\end{equation}
where $P_{A}$\ is the projection of the quantum state induced by Alice's
measurement. It coincides with the retarded projection induced by Bob's measurement.

In view of what will follow, we note that the unitary part of Alice's action
does not change the problem setting, namely the content of register $B$. This
is because $B$ is the \textit{control register}\footnote{This means that the
content of register $B$ affects the output of $U_{f}$ while remaining
unaltered through it.} of function evaluation $U_{f}$ and the unitary
transformations $H_{A}$ and $\Im_{A}$ do not apply to $B$. Correspondingly
$\Im_{A}U_{f}H_{A}$\ sends the input into the output independently term by
term keeping the value of $\mathbf{b}$\ unaltered:%
\begin{equation}
\forall\mathbf{b}:\Im_{A}U_{f}H_{A}\left\vert \mathbf{b}\right\rangle
_{B}\left\vert 00\right\rangle _{A}\left(  \left\vert 0\right\rangle
_{V}-\left\vert 1\right\rangle _{V}\right)  =\left\vert \mathbf{b}%
\right\rangle _{B}\left\vert \mathbf{b}\right\rangle _{A}\left(  \left\vert
0\right\rangle _{V}-\left\vert 1\right\rangle _{V}\right)  . \label{una}%
\end{equation}

\subsubsection{Advanced knowledge}

Let us consider the random phase representation of the reduced density
operator of register $A$ in the output state (\ref{outputa}):%
\begin{equation}
\left\vert \psi\right\rangle _{A}=\frac{1}{2}\left(  \operatorname{e}%
^{i\varphi_{0}}\left\vert 00\right\rangle _{A}+\operatorname{e}^{i\varphi_{1}%
}\left\vert 01\right\rangle _{A}+\operatorname{e}^{i\varphi_{2}}\left\vert
10\right\rangle _{A}+\operatorname{e}^{i\varphi_{3}}\left\vert 11\right\rangle
_{A}\right)  . \label{reduced}%
\end{equation}
\ The usual representation is $\rho_{A}=\left\langle \left\vert \psi
\right\rangle _{A}\left\langle \psi\right\vert _{A}\right\rangle
_{\forall\varphi_{i}}=\frac{1}{4}\left(  \left\vert 00\right\rangle
_{A}\left\langle 00\right\vert _{A}+\left\vert 01\right\rangle _{A}%
\left\langle 01\right\vert _{A}+...\right)  $. Let $\mathcal{E}_{A}$\ be the
entropy of $\left\vert \psi\right\rangle _{A}$ (or $\rho_{A}$); we have
$\mathcal{E}_{A}=n$ bit, here $n=2$.\ 

The zeroing of $\mathcal{E}_{A}$ can be due to either the projection of the
quantum state associated with the measurement of $\hat{B}$ in the initial
state (\ref{in}), retarded at the end of Alice's problem-solving action, or
that associated with the measurement of $\hat{A}$ in the output state
(\ref{outputa}). The present work is an exploration of the assumption that the
projection shares between the two measurements.

We assume that the two measurements reduce to partial measurements such that:
\textbf{1} together, they select whatever has been selected by the complete
measurements and \textbf{2} each performed alone, contribute in a
complementary and non-redundant way to the zeroing of $\mathcal{E}_{A}$. By
this we mean that no information about the solution provided by one partial
measurement is provided by the other. We should assume that the eigenvalues
selected by the two partial measurements together are those selected by the
complete measurements. We call \textbf{1} and \textbf{2} \textit{Occam
conditions}. The assumption that the two measurements contribute equally to
the zeroing of $\mathcal{E}_{A}$ explains the speedup of the present quantum algorithm.

We provide an example of reduction. We keep the assumption that the
measurement of $\hat{B}$ randomly selects the eigenvalue $10$\ and that Bob
chooses to hide the ball in drawer $01$. We assume, for simplicity of
exposition, that $U_{B}$\ bit-by-bit changes zeroes into ones and vice-versa.
As a consequence, the zeroes and ones in the outcome of the initial
measurement of $\hat{B}$\ become respectively ones and zeroes in the outcome
of the final measurement of $\hat{A}$ ($10$ becomes $01$, etc.).

Let us open a parenthesis. Here the term "selection" is used in the sense it
has in the measurement postulate, which states that the measurement of an
observable selects one of its eigenvalues, no matter the a-priori probability
of getting it. In the present context, it can be useful \ to distinguish
between the two limiting cases we are dealing with: (i) the selection of the
eigenvalue is performed with probability one of getting it and (ii) it is
performed with even probability of getting anyone of all the possible
eigenvalues. When needed for clarity, we will call the former kind of
selection "reading the eigenvalue", the latter "determining the eigenvalue".

Let $\hat{B}_{0}$ ( $\hat{B}_{1}$), of eigenvalue $b_{0}$ ($b_{1}$),\ be the
content of the left (right) cell of register $B$ and $\hat{A}_{0}$ ($\hat
{A}_{1}$), of eigenvalue $a_{0}$ ($a_{1}$), that of the left (right) cell of
register $A$ -- here $b_{0},b_{1},a_{0},a_{1}$ belong to $\left\{
0,1\right\}  $. We assume that the measurement of $\hat{B}$, with outcome
$\mathbf{b}=10$,\ reduces to, say, that of $\hat{B}_{0}$, with outcome
$b_{0}=1$ (the left digit of $10$)\ and the measurement of $\hat{A}$, with
outcome $\mathbf{b}=01$,\ to that of $\hat{A}_{1}$, with outcome $a_{1}=1$
(the right digit of $01$).

To reconstruct the selections performed by the complete measurements, we
should propagate forward in time, by $\Im_{A}U_{f}H_{A}U_{B}$, the projection
of the quantum state associated with the former measurement and backward in
time, by the inverse of $\Im_{A}U_{f}H_{A}U_{B}$, that associated with the
latter. By propagating a projection, we mean the two ends of it, namely the
states immediately before and after it. The former propagation yields that the
measurement of $\hat{A}_{0}$ in the output state selects $a_{0}=0$ (the left
digit of $01$), the latter that the measurement of $\hat{B}_{1}$ in the input
state selects $b_{1}=0$ (the right digit of $10$).

We have reconstructed the selections performed by the complete measurements by
replacing these measurements by partial measurements satisfying Occam
conditions. Note that the reduction of $\mathcal{E}_{A}$\ ascribed to each
partial measurement is one bit and that no information acquired by either
partial measurement is acquired by the other.

We show that sharing between Bob's and Alice's measurements the zeroing of
$\mathcal{E}_{A}$ does not affect Bob's freedom of choosing the number of the
drawer with the ball. In the representation of the quantum algorithm to Alice,
the probability that the measurement of $\hat{A}$\ in state (\ref{outputa})
selects $01$ (the number of the drawer with the ball chosen by Bob) is one.
The same holds for the measurement of $\hat{A}_{1}$, which selects the right
digit of $01$ with probability one. Then, applying the above said distinction,
we should say that the measurement of $\hat{A}_{1}$\ reads the right digit of
the number of the drawer with the ball $01$\ freely chosen (determined) by
Bob, without possibly altering it, or affecting Bob's freedom of choosing it.
The quantum feedback (the propagation back in time of the projection of the
quantum state due to the measurement of $\hat{A}_{1}$) does not determine any
part of Bob's choice, it determines the right digit of the random outcome of
Bob's measurement, which is before that choice.

We also note that we are not sending a message backward in time. Each of the
bits that specify the outcome of Bob's measurement is independently and
randomly selected. We are just ascribing half of these random selections to
Alice's rather than Bob's measurement.

It should be noted that this kind of\textit{ }retrocausation is sometimes
invoked to explain EPR non-locality, but mostly as a curiosity because it is
believed to have no consequences.

The measurement of $\hat{A}_{1}$ in the output state to Alice (\ref{outputa}),
which in present assumptions reads the eigenvalue $a_{1}=1$ (the right digit
of the number of the drawer with the ball $01$), projects this state on:
\begin{equation}
\left\vert \chi\right\rangle =\frac{1}{2}\left(  \operatorname{e}%
^{i\varphi_{1}}\left\vert 01\right\rangle _{B}\left\vert 01\right\rangle
_{A}+\operatorname{e}^{i\varphi_{3}}\left\vert 11\right\rangle _{B}\left\vert
11\right\rangle _{A}\right)  \left(  \left\vert 0\right\rangle _{V}-\left\vert
1\right\rangle _{V}\right)  . \label{co}%
\end{equation}

We have said that this projection must propagate backward in time through the
inverse of $\Im_{A}U_{f}H_{A}U_{B}$ until $t_{0}$, when it selects the right
digit of the random outcome of Bob's measurement $10$.\ Let us see the value
of this back-in-time propagation at time $t_{1}$ immediately after the
application of $U_{B}$ and before that of $\Im_{A}U_{f}H_{A}$. We should
advance the two ends of the projection of state (\ref{outputa}) on state
(\ref{co}) by the inverse of $\Im_{A}U_{f}H_{A}$. The result is the projection
of state (\ref{ina}),\ the input state of the quantum algorithm to Alice, on:%
\begin{equation}
H_{A}^{\dag}U_{f}^{\dag}\Im_{A}^{\dag}\left\vert \chi\right\rangle =\frac
{1}{2}\left(  \operatorname{e}^{i\varphi_{1}}\left\vert 01\right\rangle
_{B}+\operatorname{e}^{i\varphi_{3}}\left\vert 11\right\rangle _{B}\right)
\left\vert 00\right\rangle _{A}\left(  \left\vert 0\right\rangle
_{V}-\left\vert 1\right\rangle _{V}\right)  . \label{adv}%
\end{equation}

This is an outstanding consequence. State (\ref{adv}), the input state to
Alice under the assumption that the selection of the solution shares equally
between Bob's and Alice's measurements,\ tells her, before she opens any
drawer, that the number of the drawer with the ball is either $01$ or $11$ --
that the right digit of the number is $1$.

We are at a fundamental level where knowing is doing $\left[  24\right]  $.
Alice is the problem solver, her knowing in advance that the right digit of
the number of the drawer with the ball is $1$ simply means that the quantum
algorithm can locate the ball knowing it is in drawer $\mathbf{b}=01$ or
$\mathbf{b}=11$ -- ie knowing that $\mathbf{b}$ belongs to $\left\{
01,11\right\}  _{B}$, a subset of the set of all the problem settings
$\sigma_{B}$.

Of course, there are many ways of knowing in advance that the ball is in a
pair of drawers (that includes the one with the ball) and choosing the drawer
to open. It turns out that the quantum algorithm can be seen as a sum over
classical histories in each of which Alice knows in advance that the ball is
in one of two drawers and locates it by opening either drawer.

Let us see this in more detail. We should see the quantum algorithm under the
perspective of Feynman's sum over histories $\left[  25\right]  $. A history
is a classical trajectory of the quantum registers, namely a causal sequence
of sharp register states. For example:
\begin{equation}
\operatorname{e}^{i\varphi_{1}}\left\vert 01\right\rangle _{B}\left\vert
00\right\rangle _{A}\left\vert 0\right\rangle _{V}\overset{H_{A}}{\rightarrow
}\operatorname{e}^{i\varphi_{1}}\left\vert 01\right\rangle _{B}\left\vert
11\right\rangle _{A}\left\vert 0\right\rangle _{V}\overset{U_{f}}{\rightarrow
}\operatorname{e}^{i\varphi_{1}}\left\vert 01\right\rangle _{B}\left\vert
11\right\rangle _{A}\left\vert 0\right\rangle _{V}\overset{\Im_{A}%
}{\rightarrow}\operatorname{e}^{i\varphi_{1}}\left\vert 01\right\rangle
_{B}\left\vert 01\right\rangle _{A}\left\vert 0\right\rangle _{V}. \label{one}%
\end{equation}

The left-most state is one of the elements of the input\ state superposition
(\ref{ina}). The state after each arrow is one of the elements of the quantum
superposition generated by the unitary transformation of the state before the
arrow; the transformation in question is specified above the arrow.

One can check that the quantum algorithm can be seen as a sum over classical
histories in each of which Alice, given her advanced knowledge of one of the
possible halves of the problem setting, performs the function evaluations
logically necessary to find the solution.

For example, in history (\ref{one}), the problem setting is $\mathbf{b}=01$.
Alice performs function evaluation for $\mathbf{a}=11$ (second and third
state). Therefore we must assume that Alice's advanced knowledge is
$\mathbf{b}\in\left\{  01,11\right\}  _{B}$. Since the output of function
evaluation is zero (the content of register $V$\ remains unaltered), she finds
that the problem setting must be $\mathbf{b}=01$.

We must still examine Grover algorithm for $n>2$. We resort to Grover/Long
algorithm, which always provides the solution with certainty with $K=\frac
{\pi}{4\arcsin2^{-n/2}}\approx\frac{\pi}{4}2^{n/2}$ function evaluations.

With $\mathcal{R}=\frac{1}{2}$, the number of function evaluations foreseen by
the present retrocausality model is $2^{n/2}-1\approx2^{n/2}$. In fact, Alice
knows in advance $\mathcal{R}n$ bits of the number of the drawer with the
ball, namely $n/2$ bits with $\mathcal{R}=\frac{1}{2}$. If, in the worst case,
she opens $2^{n/2}-1$\ drawers without finding the ball, then she knows it is
in the only drawer she did not open.

$2^{n/2}-1$ is slightly above the number required in the optimal case. To have
$K$ exactly when $n$ goes past $2$, the value of $\mathcal{R}$ should slightly
go above $\frac{1}{2}$. In fact, Alice's advanced knowledge must slightly go
above $\frac{n}{2}$ bits to reduce the number of function evaluation by the
factor $\frac{\pi}{4}$ (with respect to the number required with the advanced
knowledge of $\frac{n}{2}$\ bits). We note that the increment of advanced
knowledge must be less than one bit -- knowing in advance an extra bit would
halve the number of function evaluations. Moreover, in the limit for
$n\rightarrow\infty$, we have again $\mathcal{R}=\frac{1}{2}$.

In any case, $\mathcal{R}=\frac{1}{2}$ \ always corresponds to an existing
quantum algorithm (Long's algorithm tuned on $2^{n/2}-1$ function evaluations)
and\ provides a good approximation of the number of function evaluations
required by the optimal one.

\subsection{Oracle quantum computing}

Given a generic oracle problem, we provide the procedure for calculating
Alice's advanced knowledge for quantum retrocausality $R=\frac{1}{2}$. This
also yields the number of function evaluations to solve the problem quantally,
which is that of a classical algorithm which benefits of the same advanced knowledge.

A generic oracle problem can be formulated as follows. We have a set of
functions $f_{\mathbf{b}}:\left\{  0,1\right\}  ^{n}\rightarrow\left\{
0,1\right\}  ^{m}$ with $m\leq n$. The suffix $\mathbf{b}$ ranges over the set
of all the problem settings $\sigma_{B}$. Bob chooses one of these functions
(a value of $\mathbf{b}$) and gives Alice the black box that computes it.
Alice, who knows the set of functions but not the one chosen by Bob, is to
find a certain feature of the function (eg the value of $\mathbf{b}$ in the
algorithm of Grover) by performing function evaluations. We call the feature
in question, which is the solution of the problem and a function of
$\mathbf{b}$, $s\left(  \mathbf{b}\right)  $. By the way, in oracle quantum
computing it is customary to call the black box \textit{oracle }and function
evaluation \textit{oracle query}.

Provided that a register $B$ contains the problem setting $\mathbf{b}$ and a
register $A$ will eventually contain the solution $s\left(  \mathbf{b}\right)
$, the most general form of the input and output states of the unitary part of
Alice's problem solving action, with respect to Alice, is:%
\begin{align}
\left\vert \operatorname{in}\right\rangle _{BAW}  &  =\frac{1}{\sqrt{c}%
}\left(
{\displaystyle\sum\limits_{\mathbf{b~\in~}\sigma_{B}}}
\operatorname{e}^{i\varphi_{\mathbf{b}}}\left\vert \mathbf{b}\right\rangle
_{B}\right)  \left\vert 00...\right\rangle _{A}\left\vert \psi\right\rangle
_{W},\label{input}\\
\left\vert \operatorname{out}\right\rangle _{BAW}  &  =U\left\vert
\operatorname{in}\right\rangle _{BAW}=\frac{1}{\sqrt{c}}%
{\displaystyle\sum\limits_{\mathbf{b~\in~}\sigma_{B}}}
\operatorname{e}^{i\varphi_{\mathbf{b}}}\left\vert \mathbf{b}\right\rangle
_{B}\left\vert s\left(  \mathbf{b}\right)  \right\rangle _{A}\left\vert
\varphi\left(  \mathbf{b}\right)  \right\rangle _{W}, \label{out}%
\end{align}
where $c$ is the cardinality of $\sigma_{B}$, $\left\vert \psi\right\rangle
_{W}$ and $\left\vert \varphi\left(  \mathbf{b}\right)  \right\rangle _{W}%
\ $are normalized states$\ $of a register $W$, which stands for any other
register or set of registers, and $U$\ is a unitary transformation.

Given that the input is stochastically defined, for $U$ to be unitary it is
not necessary that $s\left(  \mathbf{b}\right)  $ is an invertible function of
$\mathbf{b}$ like in Grover algorithm, it suffices that the values of
$s\left(  \mathbf{b}\right)  $ partition the set $\sigma_{B}$ into disjoint
blocks -- this will be the case of the quantum algorithms examined in the
following sections.

We note that the unitary part of Alice's action should not change the problem
setting, namely the content of register $B$. To this end, register $B$ should
be the control register of all function evaluations and the unitary
transformations before and after each function evaluation should not apply to
$B$. Correspondingly $U$\ sends the input into the output independently term
by term and keeping the value of $\mathbf{b}$\ unaltered:%

\begin{equation}
\forall\mathbf{b}:U\left\vert \mathbf{b}\right\rangle _{B}\left\vert
00...\right\rangle _{A}\left\vert \psi\right\rangle _{W}=\left\vert
\mathbf{b}\right\rangle _{B}\left\vert s\left(  \mathbf{b}\right)
\right\rangle _{A}\left\vert \varphi\left(  \mathbf{b}\right)  \right\rangle
_{W}. \label{indi}%
\end{equation}

Note that, for equation (\ref{indi}), the projection of the quantum state
induced by any measurement on the content of register $B$ in the output state,
advanced at time $t_{1}$ (at the time of the input state), becomes the
projection induced by performing the same measurement in the input state. This
goes along with the fact that the reduced density operator of register $B$
remains the same throughout the unitary part of Alice's action $U$. Its random
phase representation is%
\begin{equation}
\left\vert \psi\right\rangle _{B}=\frac{1}{\sqrt{c}}\left(
{\displaystyle\sum\limits_{\mathbf{b~\in~}\sigma_{B}}}
\operatorname{e}^{i\varphi_{\mathbf{b}}}\left\vert \mathbf{b}\right\rangle
_{B}\right)  \label{red}%
\end{equation}
throughout $U$. By the way, we have $\rho_{B}=\left\langle \left\vert
\psi\right\rangle _{B}\left\langle \psi\right\vert _{B}\right\rangle
_{\forall\varphi_{i}}$.

We go to the problem of computing Alice's advanced knowledge for
$\mathcal{R}=\frac{1}{2}$. For simplicity, we can assume that the problem
setting chosen by Bob is directly the outcome of the measurement of $\hat{B}$;
this saves us the distinction between input state and an initial state which
is changed into the input state by $U_{B}$. We should reduce in all the
possible ways the measurement of $\hat{B}$ in the input state and the
measurement of $\hat{A}$ in the output state to two partial measurements --
say of $\hat{B}_{i}$ and $\hat{A}_{j}$ -- such that:

\begin{enumerate}
\item[\textbf{I}] together, they select whatever has been selected by the
complete measurements and

\item[\textbf{II}] each performed alone, contribute in an equal and
non-redundant way to the selection of the solution.
\end{enumerate}

Point \textbf{II}\ implies the following two conditions. One is:%
\begin{equation}
\Delta\mathcal{E}_{A}\left(  \hat{B}_{i}\right)  =\Delta\mathcal{E}_{A}\left(
\hat{A}_{j}\right)  , \label{equg}%
\end{equation}
where $\Delta\mathcal{E}_{A}\left(  \hat{B}_{i}\right)  $ is the reduction of
$\mathcal{E}_{A}$ due to the measurement of $\hat{B}_{i}$ in the input state,
$\Delta\mathcal{E}_{A}\left(  \hat{A}_{j}\right)  $ that due to the
measurement of $\hat{A}_{j}$ in the output state. The other is:%
\begin{equation}
no~partial~measurement~outcome\ provides~enough~information~to~select~the~solution
\label{no}%
\end{equation}
In fact the cases are two: if both outcomes contained enough information, then
there would be redundant information, what is forbidden by the no-redundancy
condition. If only one did, then the two partial measurements would not
contribute equally to the selection of the solution, what is forbidden by the
equality condition (\ref{equg}).

In the case that the problem setting is an unstructured bit string, as in
Grover algorithm, condition (\ref{no}) is redundant with (\ref{equg}). It is
not when the bit string is structured.

Given a pair of measurement, of $\hat{B}_{i}$ in (\ref{input}) and $\hat
{A}_{j}$ in (\ref{out}), that satisfies conditions \textbf{I} and \textbf{II},
we obtain Alice's advanced knowledge as in Section 3.1.2. We advance by
$U^{\dag}$\ the (two ends of the) projection induced by the measurement of
$\hat{A}_{j}$ in state (\ref{out}). This projects the maximally mixed state of
register $B$ throughout $U$, namely state (\ref{red}), on a state of lower
entropy of the form%
\begin{equation}
\frac{1}{\sqrt{c^{\prime}}}\left(
{\displaystyle\sum\limits_{\mathbf{b~\in~}\sigma_{B}^{\prime}}}
\operatorname{e}^{i\varphi_{\mathbf{b}}}\left\vert \mathbf{b}\right\rangle
_{B}\right)  , \label{lower}%
\end{equation}
which represents Alice's advanced knowledge. Here $\sigma_{B}^{\prime}$ is a
subset of $\sigma_{B}$, of cardinality $c^{\prime}$; throughout $U$,\ Alice
knows in advance that $\mathbf{b}\in\sigma_{B}^{\prime}$. \ 

This way of assessing Alice's advanced knowledge requires knowing the unitary
transformation $U$ (its inverse in fact). We provide an alternative way that
bypasses this need. It suffices to note that Alice is forbidden to read
(measure) the content of register $B$ but only until the end of the unitary
part of her problem-solving action. After that, she is free to measure it.
Knowledge of the problem setting acquired after the end of the problem-solving
action does not alter the terms of the problem. Therefore, we can replace the
final Alice's measurement of the content of $A$\ in the output state
(\ref{out})\ by that of the content of $B$. This measurement also selects the
content of $A$, which in the output state is a function of that of $B$ (the
solution of the problem is a function of the problem setting).

This leaves us with two complete measurements of the content of $B$, one
performed by Bob in the input state, the other by Alice in the output state.
In order to obtain Alice's advanced knowledge, in the first place we need to
reduce -- in all the possible ways -- this pair of measurements to a pair of
partial measurements, of $\hat{B}_{i}$\ and $\hat{B}_{j}$, submitted to
conditions \textbf{I} and \textbf{II } (with $\hat{A}_{j}$\ replaced by
$\hat{B}_{j}$). Advancing at the time of the input state, by $U^{\dag}$, the
projection induced by the measurement of $\hat{B}_{j}$\ in the output state
would project the maximally mixed state of register $B$ (\ref{red}) on state
(\ref{lower}), which represents Alice's advanced knowledge.

However, this time we can avoid the advancement in question. For equation
(\ref{indi}), the projection of the quantum state induced by any measurement
on the content of register $B$ in the output state, advanced by $U^{\dag}$ at
the time of the input state, becomes the projection induced by performing the
same measurement in the input state. Therefore, we can simply measure also
$\hat{B}_{j}$\ in the input state. This directly projects the maximally mixed
state of register $B$ (\ref{red}) on state (\ref{lower}), which represents
Alice's advanced knowledge.

This latter way of assessing Alice's advanced knowledge highlights a symmetry
hidden in the former one. We are left with two partial measurements of the
content of register $B$, both performed in the input state, that satisfy
conditions \textbf{I} and \textbf{II}. We can loose the memory of which
partial measurement is performed by Alice and which by Bob. Evidently either
partial measurement can be the one performed by Alice and projects the
maximally mixed state of register $B$ on an instance of Alice's advanced
knowledge. By the way, in this sense we can say that, with $\mathcal{R}%
=\frac{1}{2}$, Alice knows "half" of the problem setting in advance.

We provide an exemplification in the four drawer case. The pairs $\hat{B}_{i}%
$\ and $\hat{B}_{j}$ whose measurements satisfy conditions \textbf{I} and
\textbf{II }are all the pairs out of the three observables $\hat{B}_{0}$,
$\hat{B}_{1}$, and $\hat{B}_{X}=\operatorname*{XOR}\left(  \hat{B}_{0},\hat
{B}_{1}\right)  $, where $\operatorname*{XOR}$ stands for \textit{exclusive
or}. With problem setting $\mathbf{b}=01$, the eigenvalues selected by the
corresponding measurements are respectively $b_{0}=0$ (the left digit of
$01$), $b_{1}=1$ (the right digit of $01$), and $\operatorname*{XOR}\left(
b_{0},b_{1}\right)  =1$ (the exclusive or between left and right digit of
$01$). As $b_{0}=0$ means $\mathbf{b}\in\left\{  01,00\right\}  _{B}$, etc.,
the corresponding instances of Alice's advanced knowledge are $\mathbf{b}%
\in\left\{  01,00\right\}  _{B}$, $\mathbf{b}\in\left\{  01,11\right\}  _{B}$,
and $\mathbf{b}\in\left\{  01,10\right\}  _{B}$, as obvious in hindsight with
problem setting $\mathbf{b}=01$\ and with $R=\frac{1}{2}$. Given a pair of
measurements, of $\hat{B}_{i}$\ and $\hat{B}_{j}$, that satisfy \textbf{I} and
\textbf{II}, for short we say that either measurement "projects" $\sigma_{B}$
on a proper subset thereof that represent Alice's advanced knowledge.

We can also see that, given the problem setting -- ie the bit string
$\mathbf{b}$\ -- any partial measurement of the content of register $B$\ is in
bijective correspondence with the subset of $\sigma_{B}$ on which it projects
$\sigma_{B}$ (see the above exemplification). Thus, summarizing:

\begin{enumerate}
\item[a)] Given a problem setting, we should find all the pairs of subsets of
$\sigma_{B}$ such that the corresponding measurements of $\hat{B}_{i}$ and
$\hat{B}_{j}$ in the input state satisfy \textbf{I} and \textbf{II} (with
$\hat{A}_{j}$\ replaced by $\hat{B}_{j}$); either subset represents a part of
the problem setting that Alice knows in advance. Note that the states of
register $W$ play no part in the calculation of the advanced knowledge. Thus
we can work with $\left\vert \operatorname{in}\right\rangle _{BA}$ and
$\left\vert \operatorname{out}\right\rangle _{BA}$ only, namely the partial
traces over $W$ of respectively $\left\vert \operatorname{in}\right\rangle
_{BAW}$ and $\left\vert \operatorname{out}\right\rangle _{BAW}$. These partial
traces, in turn, can be written solely on the basis of the oracle problem, it
suffices to know all the pairs $\mathbf{b}$ and $s\left(  \mathbf{b}\right)  $.

\item[b)] The quantum algorithm can be seen as a sum over classical histories
in each of which Alice knows in advance one of the possible halves of the
problem setting -- as exactly specified by point a) -- and performs the
function evaluations still necessary to identify the solution.
\end{enumerate}

We call a) and b) \textit{the advanced knowledge rule}. Given an oracle
problem, such that there can be a unitary transformation between the
corresponding input (\ref{input}) and output (\ref{out}), this rule defines
the number of function evaluations required for its solution with quantum
retrocausality $R=\frac{1}{2}$.

\section{Deutsch\&Jozsa algorithm}

In Deutsch\&Jozsa's $\left[  26\right]  $ problem, the set of functions is all
the constant and \textit{balanced} functions (with the same number of zeroes
and ones) $f_{\mathbf{b}}:\left\{  0,1\right\}  ^{n}\rightarrow\left\{
0,1\right\}  $. Array (\ref{dj}) gives four of the eight functions for $n=2$.%
\begin{equation}%
\begin{tabular}
[c]{|l|l|l|l|l|l|}\hline
$\mathbf{a}$ & ${\small \,f}_{0000}\left(  \mathbf{a}\right)  $ &
${\small f}_{1111}\left(  \mathbf{a}\right)  $ & ${\small f}_{0011}\left(
\mathbf{a}\right)  $ & ${\small f}_{1100}\left(  \mathbf{a}\right)  $ &
...\\\hline
00 & 0 & 1 & 0 & 1 & ...\\\hline
01 & 0 & 1 & 0 & 1 & ...\\\hline
10 & 0 & 1 & 1 & 0 & ...\\\hline
11 & 0 & 1 & 1 & 0 & ...\\\hline
\end{tabular}
\label{dj}%
\end{equation}

The bit string $\mathbf{b}\equiv b_{0}b_{1}...b_{2^{n}-1}$ is both the suffix
and the table of the function $f_{\mathbf{b}}\left(  \mathbf{a}\right)  $ --
the sequence of function values for increasing values of the argument; the
reason for labeling the function in this way will soon become clear. Alice is
to find whether the function chosen by Bob is constant or balanced by
computing $f_{\mathbf{b}}\left(  \mathbf{a}\right)  $ for appropriate values
of $\mathbf{a}$. Classically, this requires in the worst case a number of
function evaluations exponential in $n$. It requires just one function
evaluation in the quantum case.

In the following, we show that the speedup of Deutsch\&Jozsa algorithm is
explained by $\mathcal{R}=\frac{1}{2}$.

\subsection{Time-symmetric representation to Alice}

The input and output states of the quantum algorithm to Alice are
respectively:%
\[
\left\vert \psi\right\rangle =\frac{1}{4}\left(  \operatorname{e}%
^{i\varphi_{0}}\left\vert 0000\right\rangle _{B}+\operatorname{e}%
^{i\varphi_{1}}\left\vert 1111\right\rangle _{B}+\operatorname{e}%
^{i\varphi_{2}}\left\vert 0011\right\rangle _{B}+\operatorname{e}%
^{i\varphi_{3}}\left\vert 1100\right\rangle _{B}+...\right)  \left\vert
00\right\rangle _{A}\left(  \left\vert 0\right\rangle _{V}-\left\vert
1\right\rangle _{V}\right)  ,
\]%
\[
H_{A}U_{f}H_{A}\left\vert \psi\right\rangle =\frac{1}{4}\left[  \left(
\operatorname{e}^{i\varphi_{0}}\left\vert 0000\right\rangle _{B}%
-\operatorname{e}^{i\varphi_{1}}\left\vert 1111\right\rangle _{B}\right)
\left\vert 00\right\rangle _{A}+\left(  \operatorname{e}^{i\varphi_{2}%
}\left\vert 0011\right\rangle _{B}-\operatorname{e}^{i\varphi_{3}}\left\vert
1100\right\rangle _{B}\right)  \left\vert 10\right\rangle _{A}+...\right]
\left(  \left\vert 0\right\rangle _{V}-\left\vert 1\right\rangle _{V}\right)
.
\]
Registers $B$, $A$, and $V$ have the same function as in Grover algorithm,
$H_{A}$ is the Hadamard transform on register $A$ and $U_{f}$ is function evaluation.

Measuring $\hat{A}$\ in the output\ state says that the function is constant
if the measurement outcome is all zeros, balanced otherwise.

\subsection{Advanced knowledge}

We apply the advanced knowledge rule -- Section 3.2. Given the problem setting
of a balanced function, there is only one pair of partial measurements of the
content of register $B$ compatible with conditions \textbf{I} and \textbf{II}.
With, say, $\mathbf{b}$ $=0011$, $\hat{B}_{i}$ must be the content of the left
half of register $B$ and $\hat{B}_{j}$ that of the right half. The measurement
of $\hat{B}_{i}$ yields all zeros, that of $\hat{B}_{j}$ all ones.

In fact, a partial measurement yielding both zeroes and ones would violate
condition (\ref{no}): it would provide enough information to identify the
solution -- the fact that $f_{\mathbf{b}}$ is balanced. Given that either
partial measurement yields all zeroes or all ones, it must concern the content
of half register. Otherwise either equation (\ref{equg}) would be violated or
the problem setting would not be completely determined, as readily checked.\ 

One can see that, with $\mathbf{b}$ $=0011$, the measurement $\hat{B}_{i}$
performed alone selects the subset $\left\{  0011,0000\right\}  _{B}$, that of
$\hat{B}_{j}$ the subset$\ \left\{  0011,1111\right\}  _{B}$. Either subset
represents a half of the problem setting that Alice knows in
advance.\ Equation (\ref{equg}) is satisfied with $\Delta\mathcal{E}%
_{A}\left(  \hat{B}_{i}\right)  =\Delta\mathcal{E}_{A}\left(  \hat{B}%
_{j}\right)  =1$ bit.

The case of the problem setting of a constant function is analogous. The only
difference is that there are more pairs of partial measurements that satisfy
conditions \textbf{I} and \textbf{II}. Say that the problem setting is
$\mathbf{b}$ $=0000$. The measurements of the content of the left and right
half of register $B$ (each performed alone) select respectively $\left\{
0000,0011\right\}  _{B}$ and $\left\{  0000,1100\right\}  _{B}$, the
measurements of the content of even and odd cells (from the left) select
respectively $\left\{  0000,1010\right\}  _{B}$ and $\left\{
0000,0101\right\}  _{B}$, etc.

There is a shortcut to finding the subsets in question. Here the problem
setting -- the bit string $\mathbf{b}$ -- is the table of the function chosen
by Bob. For example $\mathbf{b}=0011$ is the table $f_{\mathbf{b}}\left(
00\right)  =0,f_{\mathbf{b}}\left(  01\right)  =0,f_{\mathbf{b}}\left(
10\right)  =1,f_{\mathbf{b}}\left(  11\right)  =1$. We call "good half table"
any half table in which all the values of the function are the same. One can
see that good half tables are in one-to-one correspondence with the subsets of
$\sigma_{B}$ in question. For example, the good half table $f_{\mathbf{b}%
}\left(  00\right)  =0,f_{\mathbf{b}}\left(  01\right)  =0$ corresponds to the
subset $\left\{  0011,0000\right\}  _{B}$, is the identical part of the two
bit-strings in it. Thus, given a problem setting, ie an entire table, either
good half table, or identically the corresponding subset of $\sigma_{B}$,
represents a possible instance of Alice's advanced knowledge.

Because of the structure of tables, given the advanced knowledge of a good
half table, the entire table and thus the solution can be identified by
performing just one function evaluation for any value of the argument
$\mathbf{a}$ outside the half table. Thus the advanced knowledge rule with
$R=\frac{1}{2}$ foresees that the quantum algorithm solves Deutsch\&Jozsa's
problem with certainty with just one function evaluation. This is of course in
agreement with Deutsch\&Jozsa algorithm.

A history is for example: $\operatorname{e}^{i\varphi_{2}}\left\vert
0011\right\rangle _{B}\left\vert 00\right\rangle _{A}\left\vert 0\right\rangle
_{V}\overset{H_{A}}{\rightarrow}\operatorname{e}^{i\varphi_{2}}\left\vert
0011\right\rangle _{B}\left\vert 10\right\rangle _{A}\left\vert 0\right\rangle
_{V}\overset{U_{f}}{\rightarrow}\operatorname{e}^{i\varphi_{2}}\left\vert
0011\right\rangle _{B}\left\vert 10\right\rangle _{A}\left\vert 1\right\rangle
_{V}\overset{H_{A}}{\rightarrow}\operatorname{e}^{i\varphi_{2}}\left\vert
0011\right\rangle _{B}\left\vert 10\right\rangle _{A}\left\vert 1\right\rangle
_{V}$. Since the problem setting is $\mathbf{b}=0011$ and Alice performs
function evaluation for $\mathbf{a}=10$, her advanced knowledge must be
$\mathbf{b}\in\left\{  0011,0000\right\}  _{B}$; if it were $\mathbf{b}%
\in\left\{  0011,1111\right\}  _{B}$, she would have performed function
evaluation for either $\mathbf{a}=00$ or $\mathbf{a}=01$. The result of
function evaluation, $f_{\mathbf{b}}\left(  10\right)  =1$, tells Alice that
the function chosen by Bob must be balanced.

One can see that the present analysis, like the notion of good half table,
holds unaltered for $n>2$.

\section{Simon and hidden subgroup algorithms}

In Simon's $\left[  27\right]  $ problem, the set of functions is all the
$f_{\mathbf{b}}:\left\{  0,1\right\}  ^{n}\rightarrow\left\{  0,1\right\}
^{n-1}$ such that $f_{\mathbf{b}}\left(  \mathbf{a}\right)  =f_{\mathbf{b}%
}\left(  \mathbf{c}\right)  $ if and only if $\mathbf{a}=\mathbf{c}$\ or
$\mathbf{a}=\mathbf{c}\oplus\mathbf{h}^{\left(  \mathbf{b}\right)  }$;
$\oplus$\ denotes bitwise modulo 2 addition. The bit string $\mathbf{h}%
^{\left(  \mathbf{b}\right)  }$,\textbf{ }depending on $\mathbf{b}$ and
belonging to $\left\{  0,1\right\}  ^{n}$ excluded the all zeroes string, is a
sort of period of the function.

Array (\ref{periodic}) gives four of the six functions for $n=2$. The bit
string $\mathbf{b}$ is both the suffix and the table of the function. We note
that each value of the function appears exactly twice in the table; thus 50\%
of the rows plus one always identify $\mathbf{h}^{\left(  \mathbf{b}\right)
}$.%
\begin{equation}%
\begin{tabular}
[c]{|l|l|l|l|l|l|}\hline
& ${\small h}^{\left(  0011\right)  }{\small =01}$ & ${\small h}^{\left(
1100\right)  }{\small =01}$ & ${\small h}^{\left(  0101\right)  }{\small =10}$
& ${\small h}^{\left(  1010\right)  }{\small =10}$ & ...\\\hline
$\mathbf{a}$ & ${\small f}_{0011}\left(  \mathbf{a}\right)  $ & ${\small f}%
_{1100}\left(  \mathbf{a}\right)  $ & ${\small f}_{0101}\left(  \mathbf{a}%
\right)  $ & ${\small f}_{1010}\left(  \mathbf{a}\right)  $ & ...\\\hline
00 & 0 & 1 & 0 & 1 & ...\\\hline
01 & 0 & 1 & 1 & 0 & ...\\\hline
10 & 1 & 0 & 0 & 1 & ...\\\hline
11 & 1 & 0 & 1 & 0 & ...\\\hline
\end{tabular}
\label{periodic}%
\end{equation}

Bob chooses one of these functions. Alice is to find the value of
$\mathbf{h}^{\left(  \mathbf{b}\right)  }$ by performing function
evaluation\ for appropriate values of $\mathbf{a}$.

In present knowledge, a classical algorithm requires a number of function
evaluations exponential in $n$. The quantum part of Simon algorithm solves the
hard part of this problem, namely finding a string $\mathbf{s}_{j}^{\left(
\mathbf{b}\right)  }$ "orthogonal" $\left[  27\right]  $ to $\mathbf{h}%
^{\left(  \mathbf{b}\right)  }$, with one function evaluation. There are
$2^{n-1}$ such strings. Running the quantum part of Simon algorithm yields one
of these strings at random. The quantum part is iterated until finding $n-1$
different strings. This allows Alice to find $\mathbf{h}^{\left(
\mathbf{b}\right)  }$ by solving a system of modulo 2 linear equations. Thus,
on average, finding $\mathbf{h}^{\left(  \mathbf{b}\right)  }$ requires
$\operatorname*{O}\left(  n\right)  $ iterations of the quantum part of Simon
algorithm -- in particular $\operatorname*{O}\left(  n\right)  $ function
evaluations. Moreover, if we put an upper bound to the number of iterations,
a-priori there always is a non-zero probability of not finding $n-1$ different strings.

In the following, we apply the advanced knowledge rule directly to the problem
of finding $\mathbf{h}^{\left(  \mathbf{b}\right)  }$. We do not know the
quantum algorithm that solves this problem; however, this is not necessary to
the end of finding the number of function evaluations required to solve it
with quantum retrocausality $\mathcal{R}=\frac{1}{2}$ -- see Section 3.2.
Under the assumption that $\mathcal{R}=\frac{1}{2}$ is always attainable,
Simon algorithm would turn out to be suboptimal. We will also see that the
speedup of the quantum part of Simon algorithm is explained by $\mathcal{R}%
=\frac{1}{2}$.

\subsection{Time-symmetric representation to Alice}

Knowing all the pairs $\mathbf{b}$, $s\left(  \mathbf{b}\right)
\equiv\mathbf{h}^{\left(  \mathbf{b}\right)  }$ -- from array (\ref{periodic})
-- we can write the partial trace over $W$ of the input and output states of
registers $B$ and $A$:%

\[
\left\vert \operatorname{in}\right\rangle _{BA}=\frac{1}{\sqrt{6}}\left(
\operatorname{e}^{i\varphi_{0}}\left\vert 0011\right\rangle _{B}%
+\operatorname{e}^{i\varphi_{1}}\left\vert 1100\right\rangle _{B}%
+\operatorname{e}^{i\varphi_{2}}\left\vert 0101\right\rangle _{B}%
+\operatorname{e}^{i\varphi_{3}}\left\vert 1010\right\rangle _{B}+...\right)
\left\vert 00\right\rangle _{A},
\]%
\[
\left\vert \operatorname{out}\right\rangle _{BA}=\frac{1}{\sqrt{6}}\left[
\left(  \operatorname{e}^{i\varphi_{0}}\left\vert 0011\right\rangle
_{B}+\operatorname{e}^{i\varphi_{1}}\left\vert 1100\right\rangle _{B}\right)
\left\vert 01\right\rangle _{A}+\left(  \operatorname{e}^{i\varphi_{2}%
}\left\vert 0101\right\rangle _{B}+\operatorname{e}^{i\varphi_{3}}\left\vert
1010\right\rangle _{B}\right)  \left\vert 10\right\rangle _{A}+...\right]  .
\]
Since the values of $\mathbf{h}^{\left(  \mathbf{b}\right)  }$\ partition
$\sigma_{B}$\ into disjoint blocks, there should be a unitary transformation
between the un-traced states.

\subsection{Advanced knowledge}

The analysis is similar to that of Deutsch\&Jozsa algorithm. This time a good
half table should not contain a same value of the function twice, what would
provide enough information to identify the solution of the problem (the
"period" $\mathbf{h}^{\left(  \mathbf{b}\right)  }$), thus violating condition
(\ref{no}) of the advanced knowledge rule.

This leaves us with two ways of sharing each table into two good halves. With
$\mathbf{b}=0011$, one is: $f_{\mathbf{b}}\left(  00\right)  =0,~f_{\mathbf{b}%
}\left(  10\right)  =1$ and $f_{\mathbf{b}}\left(  01\right)
=0,~f_{\mathbf{b}}\left(  11\right)  =1$; the corresponding subsets of
$\sigma_{B}$\ are $\left\{  0011,0110\right\}  _{B}$ and $\left\{
0011,1001\right\}  _{B}$; either half table or identically either subset
represents Alice's advanced knowledge of part of the problem setting. The
other is $f_{\mathbf{b}}\left(  00\right)  =0,~f_{\mathbf{b}}\left(
11\right)  =1$ and $f_{\mathbf{b}}\left(  01\right)  =0,~f_{\mathbf{b}}\left(
10\right)  =1$, etc. Equation (\ref{equg}) is always satisfied with
$\Delta\mathcal{E}_{A}\left(  \hat{B}_{i}\right)  =\Delta\mathcal{E}%
_{A}\left(  \hat{B}_{j}\right)  =\varepsilon_{A}=0.585$ bit (entropy reduction
from $-\log_{2}\frac{1}{3}$ bit to $1$ bit).

We note parenthetically that sharing each table into two halves is accidental
to Deutsch\&Jozsa's and Simon algorithms. In the quantum part of Shor's
$\left[  28\right]  $ factorization algorithm (finding the period of a
periodic function), taking two shares of the table that do not contain a same
value of the function twice implies that each share is less than half table if
the domain of the function spans more than two periods.

Given the advanced knowledge of a good half table, the entire table and then
$\mathbf{h}^{\left(  \mathbf{b}\right)  }$ can always be identified by
performing just one function evaluation for any value of the argument
$\mathbf{a}$ outside the half table. Thus, under the assumption that
$\mathcal{R}=\frac{1}{2}$ is always attainable, the advanced knowledge rule
foresees that Simon's problem can be solved with certainty with just one
function evaluation. Simon's algorithm, which requires $\operatorname*{O}%
\left(  n\right)  $ function evaluations and does not provide the solution
with certainty with a prefixed number of function evaluations, is suboptimal
under this assumption.

By the way, the fact that, with $\mathcal{R}=\frac{1}{2}$, just one function
evaluation identifies $\mathbf{h}^{\left(  \mathbf{b}\right)  }$implies that
just one function evaluation identifies all the $\mathbf{s}_{j}^{\left(
\mathbf{b}\right)  }$ orthogonal to $\mathbf{h}^{\left(  \mathbf{b}\right)  }%
$. This means that the speedup of the quantum part of Simon algorithm is
explained by $\mathcal{R}=\frac{1}{2}$.

We give the simplest instance, $n=2$, of the quantum algorithm that directly
finds $\mathbf{h}^{\left(  \mathbf{b}\right)  }$ with just one function
evaluation. We have%
\[
\left\vert \operatorname{in}\right\rangle _{BAV}=\frac{1}{2\sqrt{3}}\left(
\operatorname{e}^{i\varphi_{0}}\left\vert 0011\right\rangle _{B}%
+\operatorname{e}^{i\varphi_{1}}\left\vert 1100\right\rangle _{B}%
+\operatorname{e}^{i\varphi_{2}}\left\vert 0101\right\rangle _{B}%
+\operatorname{e}^{i\varphi_{3}}\left\vert 1010\right\rangle _{B}+...\right)
\left\vert 00\right\rangle _{A}(\left\vert 0\right\rangle _{V}-\left\vert
1\right\rangle _{V}),
\]%
\[
\left\vert \operatorname{out}\right\rangle _{BAV}=\mathcal{P}_{A}H_{A}%
U_{f}H_{A}\left\vert \operatorname{in}\right\rangle _{BAV}=\frac{1}{2\sqrt{3}%
}\left[
\begin{array}
[c]{c}%
\left(  \operatorname{e}^{i\varphi_{0}}\left\vert 0011\right\rangle
_{B}+\operatorname{e}^{i\varphi_{1}}\left\vert 1100\right\rangle _{B}\right)
\left\vert 01\right\rangle _{A}+\\
\left(  \operatorname{e}^{i\varphi_{2}}\left\vert 0101\right\rangle
_{B}+\operatorname{e}^{i\varphi_{3}}\left\vert 1010\right\rangle _{B}\right)
\left\vert 10\right\rangle _{A}+...
\end{array}
\right]  (\left\vert 0\right\rangle _{V}-\left\vert 1\right\rangle _{V}).
\]
Register $W$ reduces to the usual register $V$, meant to contain the result of
function evaluation modulo 2 added to its previous content.\ $H_{A}$ is
Hadamard on register $A$, $U_{f}$ function evaluation, $\mathcal{P}_{A}$ the
permutation of the basis vectors $\left\vert 01\right\rangle _{A}$ and
$\left\vert 10\right\rangle _{A}$. Checking whether there is the similar
algorithm for $n>2$ should be the object of further work.

The sum over histories representation can be developed as in Deutsch\&Jozsa
algorithm. If, for example, Alice's advanced knowledge is $\mathbf{b}%
\in\left\{  0011,0110\right\}  _{B}$, she can identify the value of
$\mathbf{h}^{\left(  \mathbf{b}\right)  }$ by performing a single function
evaluation for either $\mathbf{a}=01$ or $\mathbf{a}=11$ -- see array
(\ref{periodic}) -- etc.

The fact that Alice knows in advance a good half table and that, as a
consequence, she can classically identify the entire table and thus the
solution with just one function evaluation clearly holds unaltered for $n>2$.
It should also apply to the generalized Simon's problem and to the Abelian
hidden subgroup problem. In fact the corresponding algorithms are essentially
Simon algorithm. In the hidden subgroup problem, the set of functions
$f_{\mathbf{b}}:G\rightarrow W$ map a group $G$ to some finite set $W$\ with
the property that there exists some subgroup $S\leq G$ such that for any
$\mathbf{a},\mathbf{c}\in G$, $f_{\mathbf{b}}\left(  \mathbf{a}\right)
=f_{\mathbf{b}}\left(  \mathbf{c}\right)  $ if and only if $\mathbf{a}%
+S=\mathbf{c}+S$. The problem is to find the hidden subgroup $S$ by computing
$f_{\mathbf{b}}\left(  \mathbf{a}\right)  $ for the appropriate values of
$\mathbf{a}$. Now, a large variety of problems solvable with a quantum speedup
can be re-formulated in terms of the hidden subgroup problem $\left[
29,30\right]  $. Among these we find: the seminal Deutsch's problem, finding
orders, finding the period of a function (thus the problem solved by the
quantum part of Shor's\ factorization algorithm), discrete logarithms in any
group, hidden linear functions, self shift equivalent polynomials, Abelian
stabilizer problem, graph automorphism problem.

\section{Conclusion}

We have extended the representation of the quantum algorithm, conventionally
limited to the process of solving the problem, to the process of setting the
problem. The initial measurement performed by Bob (the problem setter), in a
mixture of all the possible problem settings, selects a problem setting at
random. Bob prepares the input state of the quantum algorithm by unitarily
transforming this random outcome into the desired problem setting.

This representations is with respect to Bob and any external observer, it
cannot be with respect to Alice (the problem solver), to whom the problem
setting must be hidden inside a black box. To Alice, the projection of the
quantum state induced by Bob's measurement must be retarded at the end of the
unitary part of her problem solving action. The input state to Alice remains a
mixture of all the possible problem settings. Its entropy represents Alice's
complete ignorance of the problem setting chosen by Bob. Alice, by performing
function evaluations, unitarily transforms this mixture into a mixture of
tensor products, each a problem setting and the corresponding solution. Then
she reads the solution by a final measurement.

This relational representation of the quantum algorithm is time-symmetric. We
show that, mathematically, one can ascribe any part of the selection of the
random outcome of the initial Bob's measurement to the final Alice's
measurement. This projects the input state to Alice on a state of lower
entropy where she knows the same part of the problem setting in advance,
before performing any function evaluation. She uses this information to reach
the solution with fewer function evaluations. The quantum algorithm is a sum
over classical histories in each of which Alice knows in advance part of the
problem setting and performs the function evaluations still required to find
the solution.

Given an oracle problem and a value of $\mathcal{R}$, which can be seen as the
fraction of the information that specifies the random outcome of the initial
Bob's measurement whose selection is ascribed to the final Alice's
measurement, the present retrocausality model yields the number of function
evaluations required to solve it quantally. Conversely, given a known quantum
algorithm, it yields the value of $\mathcal{R}$ that explains its speedup.

We have compared the model with the historical quantum algorithms. We have
$\mathcal{R}=\frac{1}{2}$ in all the quantum algorithms requiring a single
function evaluation and slightly above $\frac{1}{2}$\ in Grover algorithm when
more than one function evaluation is required. Conversely, $\mathcal{R}%
=\frac{1}{2}$ always yields the number of function evaluations required by an
existing quantum algorithm and a good approximation of that required by the
optimal one. If this held in general, it would solve the quantum query
complexity problem: given an oracle problem, find the order of magnitude of
the number of function evaluations (oracle queries) required to solve it
quantally in an optimal way.

This work is an exploration. Future work should compare the present model with
other classes of quantum algorithms, check whether there is the quantum
algorithm that solves Simon and the Abelian hidden subgroup problems with the
number of function evaluations foreseen by the model for $\mathcal{R}=\frac
{1}{2}$, investigate whether, given an oracle problem, $\mathcal{R}=\frac
{1}{2}$ is always attainable and the reason it should yield a good
approximation of the number of function evaluations required by the optimal
quantum algorithm, investigate what is the maximum value of $\mathcal{R}$ attainable.

Because of the fundamental character of quantum search in an unstructured
database, we conjecture that $\mathcal{R}=\frac{1}{2}$\ is always attainable
and that the maximum value of $\mathcal{R}$ physically possible is attained in
Grover algorithm.

Other foundational questions that might deserve investigation are: whether the
relation between the present form of quantum retrocausality and quantum
efficiency is confined to the complex realm of quantum computation or could
manifest itself in more elementary situations, whether there is a relation
between $\mathcal{R}$ and the temporal Bell inequalities.

\subsection*{Acknowledgments}

$\ \ $Thanks are due to David Finkelstein for useful discussions.

\subsection*{References}

$\ \ \ \left[  1\right]  $ Grover, L. K.: A fast quantum mechanical algorithm
for database search. Proc. of the 28th Annual ACM Symposium on the Theory of
Computing, 212-219 (ACM press New York, 1996)

$\left[  2\right]  $ Leggett, A. J., Garg, A.: Quantum mechanics versus
macroscopic realism: is the flux there when nobody looks?. Phys. Rev. Lett.
54, 857-860 (1985)

$\left[  3\right]  $ Braunstein, S. L., Caves, C. M.: Information theoretic
Bell inequalities. Phys. Rev. Lett. 61, 662-665 (1988)

$\left[  4\right]  $ Morikoshi, F.: Information-theoretic temporal Bell
inequality and quantum computation., Phys. Rev. A 73, 052308- 052312 (2006)

$\left[  5\right]  $ Finkelstein, D. R.: Space-time structure in high energy
interactions. In Gudehus, T., Kaiser, G., Perlmutter, A. (ed.), Fundamental
Interactions at High Energy. New York: Gordon \& Breach 324-338 (1969)

$\left[  6\right]  $ Bennett, C. H.: The Thermodynamics of Computation -- a
Review. Int. J. Theor. Phys. 21, 905-940 (1982)

$\left[  7\right]  $ Fredkin, E., Toffoli, T.: Conservative Logic. Int. J.
Theor. Phys. 21, 219-253 (1982)

$\left[  8\right]  $ Feynman, R.: Simulating Physics with Computers.
International Journal of Theoretical Physics 21 (6--7): 467-488 (1982)

$\left[  9\right]  $ Deutsch, D.: Quantum Theory, the Church Turing Principle
and the Universal Quantum Computer. Proc. Roy. Soc. London A 400, 97-117 (1985)

$\left[  10\right]  $\ Aharonov, Y., Bergman, P. G., Lebowitz, J. L.: Time
Symmetry in the Quantum Process of Measurement. Phys. Rev. B 134,\textbf{
}1410-1416 (1964)

$\left[  11\right]  $ Aharonov, Y., Albert, D., Vaidman, L.: How the result of
a measurement of a component of the spin of a spin-1/2 particle can turn out
to be 100". Physical Review Letters 60 (14), 1351-1354 (1988)

$\left[  12\right]  $ Aharonov, Y., Popescu, S.\ Tollaksen, J.: A
time-symmetric formulation of quantum mechanics. Physics today, November
issue, 27-32 (2010)

$\left[  13\right]  $\ Rovelli C.: Relational Quantum Mechanics. Int. J.
Theor. Phys. 35, 637-658 (1996)

$\left[  14\right]  $ Hawking, S.: On the Shoulders of Giants (Running Press,
Philadelphia-London 2003)$.$

$\left[  15\right]  $ Long, G.L.: Grover algorithm with zero theoretical
failure rate. Phys. Rev. A 64, 022307 (2001)

$\left[  16\right]  $ Toyama, F. M., van Dijk, W., Nogami, Y.: Quantum search
with certainty based on modified Grover algorithms: optimum choice of
parameters. Quantum Inf Process, 12, 1897-1914 (2013)

$\left[  17\right]  $ Bennett C. H., Bernstein E., Brassard G., Vazirani, U.:
Strengths and weaknesses of quantum computing. SIAM Journal on Computing 26.5,
1510-1523 (1997)

$\left[  18\right]  $ Castagnoli, G., Finkelstein, D. R.: Theory of the
quantum speedup.\textit{ }Proc. Roy. Soc. Lond. A 457, 1799-1807 (2001)

$\left[  19\right]  $\ Castagnoli, G.: The quantum correlation between the
selection of the problem and that of the solution sheds light on the mechanism
of the quantum speed up. Phys. Rev. A 82, 052334-052342 (2010)

$\left[  20\right]  $ Castagnoli, G.: Probing the mechanism of the quantum
speed-up by time-symmetric quantum mechanics. Proc. of the 92nd\ Annual
Meeting of the AAAS Pacific Division, Quantum Retrocausation: Theory and
Experiment (2011)

$\left[  21\right]  $ Morikoshi, F.: Problem-Solution Symmetry in Grover's
Quantum Search Algorithm. Int. J. Theor. Phys. 50, 1858-1867 (2011)

$\left[  22\right]  $\ Dolev S., Elitzur, A. C.: Non-sequential behavior of
the wave function. arXiv:quant-ph/0102109 v1 (2001)$\ $

$\left[  23\right]  $ Bohm, D., Pines, D. A.: Collective Description of
Electron Interactions: III. Coulomb Interactions in a Degenerate Electron Gas.
Phys. Rev. 92, 626-636 (1953)

$\left[  24\right]  $ Finkelstein, D. R., private communication

$\left[  25\right]  $ Feynman, R., Hibbs, A. R.: Quantum Mechanics And Path
Integrals (New York, McGraw-Hill 1965)

$\left[  26\right]  $\ Deutsch, D., Jozsa, R.: Rapid Solution of Problems by
Quantum Computation. Proc. R. Soc. Lond. A 439, 553-558 (1992)

$\left[  27\right]  $ Simon, D.: On the power of quantum computation. Proc.
35th Annual IEEE Symposium on the Foundations of Computer Science, 116-123 (1994)

$\left[  28\right]  $ Shor, P. W.: Algorithms for quantum computation:
discrete logarithms and factoring. Proc. 35th Annual IEEE Symposium on the
Foundations of Computer Science, 124-134 (1994)

$\left[  29\right]  $ Mosca, M., Ekert, A.: The Hidden Subgroup Problem and
Eigenvalue Estimation on a Quantum Computer. Proc. QCQC '98, selected papers
from the First NASA International Conference on Quantum Computing and Quantum
Communications, Springer-Verlag London, UK 1998,174-188 (1998)

(Springer-Verlag London, UK, 1998)

$\left[  30\right]  $ Kaye, P., Laflamme, R., Mosca, M.: An Introduction To
Quantum Computing. 146-147 (Oxford University Press, 2007)\bigskip

\end{document}